# Lamellar $L_\alpha$ mesophases doped with inorganic nanoparticles

Doru Constantin*[a] and Patrick Davidson[a]

A flourishing research activity, bringing together chemistry, physics and materials science, aims to develop nanostructured hybrid systems composed of nanoparticles with interesting properties (optical, magnetic, catalytic, etc.) dispersed within an organic matrix. In this context, a very important goal is that of controlling at the same time the position and the orientation of the particles, in a precise and reproducible way. The use of lyotropic liquid crystals as host phases is a very promising strategy that has prompted sustained experimental work over the last decade. We briefly review this field, with an accent on the structure and the physical characterization of these novel materials.

## Introduction

Elaborating and investigating hybrid materials by inserting inorganic (metallic or mineral) nanoparticles into liquid crystal phases has recently become the focus of intense worldwide research, motivated by the perspective of industrial applications as diverse as display technology, drug delivery, or information storage. Indeed, such hybrid materials may combine the typical electronic properties of inorganic materials (magnetism, light-absorption, electrical conductivity) with the characteristic properties of liquid-crystals (fluidity, anisotropy, processability, spatial confinement). From a more fundamental point of view, these hybrid materials can also help addressing new issues about novel mesophases and about how the lamellar mesophase is affected by the nanometric inclusions and how the latter experience new membrane-mediated interactions. Moreover, such organic-inorganic lamellar phases provide good model systems to mimic peptidic inclusions in biological membranes. Both these applied and fundamental considerations drive the fast expansion of this field, making this Minireview timely.

The scope of this Minireview, however, has been restricted in order to keep it to a reasonable size. For example, we will not address here the cases of either large particles (with all dimensions larger than about 100 nm) or thermotropic liquid crystals. The latter have been the subject of recent reviews.[1] Moreover, although hybrid materials obtained by inserting mineral or metallic nanoparticles into the lamellar phase of block-copolymers have recently raised much attention, this large research field will not be covered as fairly recent reviews have been dedicated to this topic.[1e, 2] Also, we will not describe the insertion of polymer coils into lamellar phases. Moreover, we will not review here any work about the DNA-surfactant complexes that have raised much interest for applications in transfection since the mid-1990s.[3] We only consider bulk materials and thus we omit thin films, Langmuir monolayers and layer-by-layer assemblies. We also do not consider here the hybrid lamellar phases of organometallic lyotropic and thermotropic compounds that display homogeneous inorganic sublayers rather than individual nanoparticles.[4] In spite of the recent reports of liquid-crystalline properties of graphene and graphene oxide suspensions[5] and of the insertion of graphene in thermotropic liquid crystals,[6] we are not yet aware of any work describing lyotropic lamellar phases doped with such carbon-based nanosheets. Most importantly, we do not review here the fast-expanding field of the synthesis of nanoparticles within liquid-crystalline templates.[7]

In recent years, the study of nanoparticles inserted within lyotropic mesophases has made remarkable progress. This development was made possible by advances in the synthesis of nanoparticles with wide-ranging sizes, morphologies and surface properties. Research in this area is motivated both by the possible applications expected, from delivery vectors for biological studies to metamaterials, and by fundamental questions, such as the coupling mechanisms between inclusions and the host mesophase. The latter aspect has prompted theoretical developments, mainly based on continuum smectic elasticity.[8]

These liquid crystalline hybrids combine microscopic-scale heterogeneity (which confines and orients the inclusions) and macroscopic-scale positional and orientational order. So far, most studies have been done on disordered samples, consisting of many small and randomly oriented domains. Although this type of analysis can provide very useful information, the use of oriented samples is essential for distinguishing (e.g. in a scattering experiment) between the symmetry axis of the phase and perpendicular directions and for taking full advantage of the anisotropic properties (optical, magnetic, dynamic etc.) of the material, which are essential for the envisioned applications. Macroscopic-scale alignment can be obtained by applying an external field or by thermal annealing (in the latter case, the alignment direction is imposed by the sample boundaries).

According to the location of particles within the phase, one can distinguish between hydrophilic inclusions (free to diffuse in the water domain or associated to the surfactant heads) and hydrophobic ones (confined in the medium of the alkyl chains). Although the relevant concepts and experimental methods involved are the same for both particle types, in the second case there is a very important practical restriction in the achievable thickness: indeed, while the water layer can be hundreds of nanometers thick, the hydrophobic medium of the chains is more difficult to swell.

[a] Dr D. Constantin and Dr P. Davidson
Laboratoire de Physique des Solides, UMR 8502 CNRS, Université Paris Sud, Bât. 510, 91405, Orsay cedex, France
Fax: (+33 1 69 15 60 86)
E-mail: doru.constantin@u-psud.fr

Among doped mesophases, the lamellar phase was by far the most studied. Irrespective of the inclusions' location within the matrix, the latter is expected to modify their interaction significantly. In particular, the confinement can reduce the effective dimension of the system and, for anisotropic particles, enhance their orientation. One can then have a coupling between the two types of order: three-dimensional smectic (for the matrix) and two-dimensional nematic or smectic (for the inclusions).

A representative example is that of "sliding phases", discovered in mixtures of DNA and cationic lipids.[9] The resulting two-dimensional smectic structure was studied in detail.[10] It is however difficult to control the structure of these systems at the mesoscopic (bilayer spacing, persistence length of the polymer) or macroscopic scales, rendering a systematic study quite difficult. At the same time, the presence of cytotoxic cationic lipids is a major drawback in therapeutic applications. A more promising strategy uses neutral lipids.[11] Possible applications of DNA/lipid phases to nano-structuring have also been considered.[12]

We will not describe in detail the experimental techniques used to study doped phases because these techniques are essentially those already used in the field of lyotropic liquid crystals, namely phase diagram identification, polarized-light microscopy, small-angle X-ray scattering, and rheology or in the field of nanoparticles, namely electron microscopy and IR-Vis-UV spectroscopy. Let us merely stress that inserting mineral or metallic nanoparticles into lamellar mesophases greatly enhances the X-ray scattering contrast and also sometimes the birefringence, which is quite beneficial from the instrumental point of view.

## Statistical physics and structure

We present the systems according to particle dimensionality and to the environment of the inclusion (in water – where the electrostatic interactions are dominant or within the membrane – where steric, Van der Waals and elastic interactions play an important role)

**Charged spherical particles in water:**

The overwhelming majority of colloidal solutions consist of charged nanospheres dispersed in water. The two most common model systems are silica and latex particles: the former can be easily obtained, with various sizes and reasonable monodispersity, while the latter can be functionalized with a wide range of surface ligands. Quite naturally, they were also among the first to be inserted within mesophases.

*Silica*
Silica spheres, 26 nm in diameter, were dispersed (up to a concentration of 0.8 vol%) in a dilute and soft lamellar phase of nonionic surfactant ($C_{12}EO_5$) and co-surfactant (hexanol).[13] The presence of the particles reduces the degree of order and the stability of the lamellar phase, features that are well captured by a thermodynamical model that includes the particles' confinement between the bilayers and their role in decreasing the steric repulsive force in the membrane stack.

Well-aligned samples of the same system (with up to 3 vol% of silica nanoparticles) were studied by SAXS.[14] The data can be explained by steric repulsion, with an effective hard-core diameter of 34 nm. The increase with respect to the particle diameter might be due to the electrostatic interaction and/or to an elastic repulsion induced by the lamellar phase.

More recently, smaller silica spheres (about 8 nm in diameter) were added to concentrated phases of the monolinolein/water system.[15] Aside from the changes in transition temperatures and in the lattice parameters of the various phases, this work also reveals a structure peak due to the interaction between particles. The authors conclude that, above a threshold concentration, the particles phase-separate out of the lipid matrix.

The particle size can play a subtle role, as shown for spheres dispersed in a soft lamellar phase of nonionic surfactant:[16] particles larger than 15 nm in diameter are covered by a surfactant bilayer, while smaller ones lack this protection and aggregate irreversibly.

*Latex*
From structural studies of the dilute lamellar phase of $C_{12}EO_5$, Imai et al.[17] argue that the presence of spherical nanoparticles enhances the steric repulsion between membranes, in contrast with the findings of Salamat et al for silica.[13]

Latex particles were also used to probe the relaxation of the host phase (see section **Microrheology and electrophoresis** below) and to control the smectic order in highly swollen $C_{12}EO_5$ lamellar systems (Figure 1).[18]

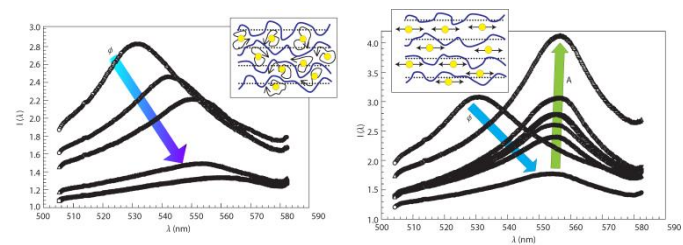

Figure 1: Left: Effect of particle doping on the smectic order. The Bragg peak intensity decreases and the peak wavelength increases with the particle concentration $\phi$. The inset schematically shows the effect of particle doping on the membrane smectic order. Right: Effects of the frequency *f* of the electric field on the latex-doped hyperswollen lamellar phase. With a decrease in *f*, the amplitude of the particle motion *A* increases and thus the Bragg peak intensity monotonically increases. The peak wavelength is insensitive to the electric field frequency. Adapted by permission from Macmillan Publishers Ltd: Nature Materials, vol. 4, pages 75-80, copyright 2004 (reference [18]).

*Polyoxometalates*
Polyoxometalates (POMs) are complex anions that are increasingly popular because of their applications in various fields such as catalysis and electrochemistry.[19] POMs have some intrinsic advantages such as their well-defined and elegant molecular structures, their inherent monodispersity, and their spectroscopic properties that make them very good candidates for insertion in lyotropic lamellar phases. One of the simplest POMs, $H_3PW_{12}O_{40}$, is commercially available and can be easily dissolved in water, yielding acidic solutions of negatively charged spherical $PW_{12}O_{40}^{3-}$ anions with diameter $D \sim 1$ nm.

The $L_\alpha$ phase of the Brij30 non-ionic industrial surfactant (essentially $C_{12}EO_4$) can be doped with large amounts $PW_{12}O_{40}^{3-}$ anions.[20] This surfactant was selected because it has no electrical charge and therefore no electrostatic interaction between anion and membrane was expected. The phase diagram of this system, represented as a function of POM doping and surfactant concentration, displays a large region of stability of the doped lamellar phase. The smectic repeat distance reaches values up to 12 nm and further dilution leads to the expulsion of excess solvent from the lamellar phase. At membrane volume fractions, $\phi_{mem}$, larger than 70%, the aqueous space between the membranes is too narrow for the POMs to be incorporated into the lamellar stack. This fairly intuitive condition that the space between membranes should be large enough to accommodate the dopant is found in many similar systems. At $\phi_{mem} = 50\%$, the POM volume fraction, $\phi_{POM}$, in the aqueous medium can reach up to 7% (i.e. ~ 50 w% mass fraction).

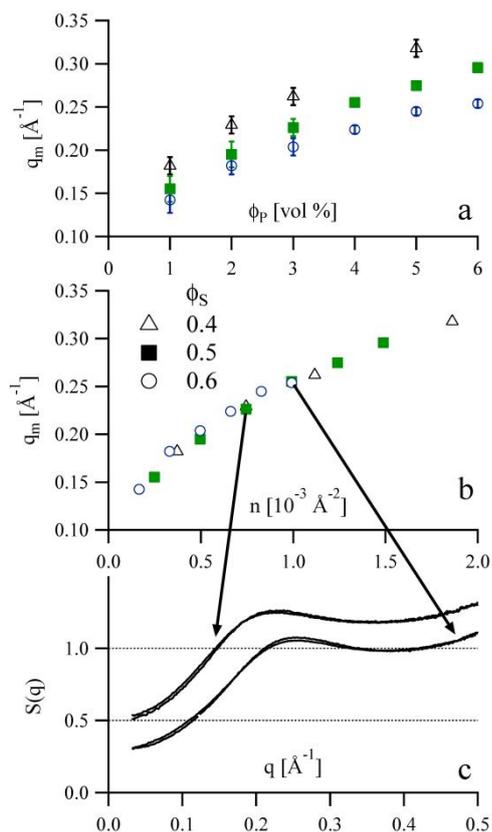

Figure 2: Position of the structure factor maximum $q_m$, plotted as a function of the volume fraction of particles $\phi_P$ (a) and as a function of their in-plane density $n$ (b) for different surfactant concentrations $\phi_S$ (various symbols and colors). In (c) we compare structure factors with the same $n$, but different $\phi_P$ and $\phi_S$. The two curves corresponding to $n = 0.75\ 10^{-3}\ Å^{-2}$, with $\phi_S = 0.4$ and 0.5, are shifted upwards by 0.2. Reprinted with permission from reference [21].

An X-ray diffraction study of the doped $L_\alpha$ phase gave clear evidence that the POMs are really incorporated between the membranes. The X-ray diffractograms show several equidistant peaks that correspond to the smectic reflections, proving the lamellar symmetry of the mesophase. Moreover, the relative intensities of the different peaks vary upon POM doping as the intensity of the second order peak strongly increases compared to that of the first one. These relative intensities directly reflect, via a Fourier transform, the content of the unit cell, which is here the smectic layer. In other words, the electronic density profile along the normal to the layers can be reconstructed from the intensities of the diffractions peaks. Close inspection of these profiles, for various surfactant and POM volume fractions, shows that, unexpectedly, the POM anions adsorb onto the polyethylene glycol (PEG) brushes that cover the surfactant membranes. It was argued that POM adsorption could be mediated by $H^+$ cations that are attracted by the ethylene oxide groups. In turn, the $H^+$ cations could attract the POMs to the membranes.

Because homeotropic samples in flat glass capillaries are readily obtained by thermal annealing, the POM-doped $L_\alpha$ phase is a very good model system to study the interaction of charged nanoparticles at interfaces by scattering techniques.[21] By shining an X-ray beam parallel to the normal to the smectic layers, the in-plane scattering of the inclusions can be recorded. Moreover, by dividing this signal by the POM form factor that is perfectly known, the 2D liquid-like structure factor of the charged particles can be retrieved and studied as a function of confinement and particle volume fraction $\phi_{POM}$. In fact, since the POMs are adsorbed on the membranes, the relevant parameter is the POM surface density on the membranes, $n$ (Figure 2).

By using statistical physics methods from the theory of liquids, it is possible to fit these 2-D structure factors and therefore determine the 2-D interaction potential between the particles. This potential includes a DLVO component (measured on an aqueous POM solution) and an additional repulsive dipolar interaction. This latter interaction of the particles at the interface arises from the difference in ionic strength between the aqueous medium and the bilayer, so that each particle acquires a dipole moment, parallel to the layer normal, and therefore experiences an additional power-law repulsion. Note, by the way, that no like-charge attractions between the charged particles at interfaces were detected in these experiments.

*Noble metals*
Noble metal particles are natural candidates for inclusion within mesophases due to their exceptional optical properties (see section **Electronic properties** below). Both silver and gold nanoparticles were confined in lamellar and hexagonal phases. To our knowledge, the first such study was made by Wang et al.,[22] who doped the lamellar phase of the SDS/n-hexanol/water/n-dodecane system with hydrophilic or hydrophobic silver spheres, about 5 nm in diameter. The dispersion state of the particles was monitored using UV-Vis spectroscopy. Using a similar approach, Firestone et al. monitored the position of the particles in the lamellar phase by the shift of the surface plasmon resonance.[23] Other preparation methods include synthesizing the particles directly within the lamellar phase[24] and dispersing the particles in a dilute micellar phase and then concentrating it by the addition of a polymer that competes with the surfactant for the hydration water.[25]

We should also mention here (without going into details) the inclusion of noble metal particles in hexagonal[26] or sponge phases.[27]

*Quantum Dots*
Semiconductor nanocrystals, or "quantum dots" (QDs) also have interesting electronic and optical properties, which are very sensitive to their size. Water-soluble QDs were incorporated in well-oriented lipid lamellar phases deposited on a solid substrate.[28] X-ray reflectivity and grazing-incidence diffraction showed that the particles were confined within the water layer and that they exhibited some degree of order within the plane. Bulk phases were also formulated.[29] Marked three-dimensional order (Figure 3) was obtained in a more complex phase, where the QDs were confined within a template formed by lipid bilayers and actin filaments.[30] Interestingly, the emission spectrum of the confined QDs exhibits a red shift that increases with their degree of order. The authors explain this feature by a superradiance effect.

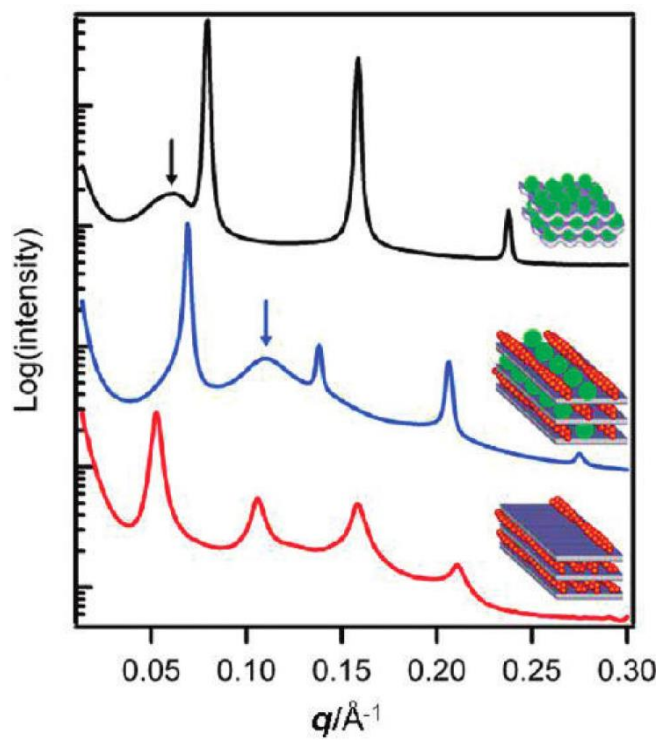

Figure 3: Powder diffraction spectra for three types of complexes: quantum dots/lipids (top), quantum dots/actin/lipids (middle) and actin/lipids (bottom). Reprinted with permission from Nano Letters, vol. 11, pages 5443-5448. Copyright (2011), American Chemical Society. (reference [30]).

**Hydrophobic spherical particles**

*Iron oxides*

One of the first examples of doped lamellar phases are the so-called "ferrosmectics" obtained by swelling a lyotropic $L_\alpha$ phase with an oil-based ferrofluid.[31] This latter component is a colloidal suspension in cyclohexane of maghemite ($\gamma$-$Fe_2O_3$) spherical ferromagnetic particles covered with surfactant molecules. These particles are polydisperse and their diameter follows a log-normal distribution with average diameter $<D> \sim 10$ nm, including the organic coating, and polydispersity $\sigma \sim 0.4$. They are small enough to be considered magnetic monodomains but large enough that their magnetic moment is locked with respect to the crystalline structure. Therefore, ferrofluids have a superparamagnetic behavior, with a linear dependence of the magnetization on field intensity at low field. The magnetization at saturation of the particles is $3.8 \times 10^5$ A.m$^{-1}$.

The host $L_\alpha$ phase is a well-known quaternary mixture of sodium dodecylsulfate (SDS) surfactant, pentanol, water, and cyclohexane.[32] The addition of pentanol as a cosurfactant decreases the bending constant, $\kappa$, of the surfactant-cosurfactant membranes, down to $kT$. Consequently, the $L_\alpha$ phase can be largely swollen with cyclohexane thanks to stabilizing repulsions between fluctuating membranes, as described by Helfrich.[33] The lamellar period, $d$, can thus reach values of the order of 40 nm, which is much larger than with usual lyotropic phases such as the $C_{12}EO_4$ system described above and also much larger than the particle diameter.

Upon swelling with the ferrofluid oily phase, a doped lamellar phase can be obtained with particle volume fraction, $\phi$, typically ranging from 0 to $\sim 5\%$, without any particle aggregation. Compared to usual $L_\alpha$ phases, this doped phase displays outstanding magnetic properties. For example, observations in polarized-light microscopy show that, when submitted to a magnetic field, the ferrosmectic orients with the layers parallel to the field at field intensities as low as $\sim 5$ mT compared to the $\sim 10$ T usually required.[31b] This proves the existence of a strong coupling between the magnetic particles and the surfactant membranes but the nature of this coupling is not obvious. Altogether, ferrosmectics combine the large magnetic susceptibility of a ferrofluid with the uniaxial and 1D-periodic symmetry of a $L_\alpha$ phase. They can also be regarded as a 1D stack of 2D magnetic liquids.

Small-angle X-ray (SAXS) and neutron (SANS) scattering are two classical techniques used to study the structure of lyotropic lamellar phases. In the specific case of ferrosmectics, SANS has two advantages over SAXS. On the one hand, the SAXS is dominated by the signal (and absorption) of the iron oxide particles, which makes it more difficult to obtain information about the surfactant membranes. On the other hand, by using 51% deuterated solvent, the particles can be almost perfectly contrast-matched for SANS, so that the scattering only comes from the surfactant membranes.[34] Note, by the way, that the magnetic contribution to the maghemite particle scattering is not negligible and must be taken into account.

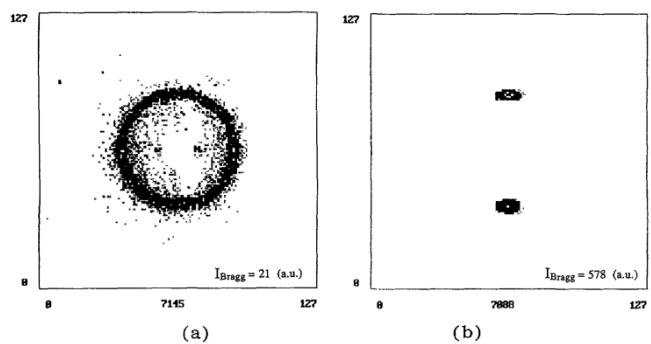

Figure 4: Two-dimensional small-angle neutron scattering spectra of a ferrosmectic phase. (a) powder sample quenched from the isotropic state. (b) oriented sample obtained by applying a rotating magnetic field. Reprinted with permission from reference [34].

Single-domain samples of ferrosmectics in planar orientation, held in square capillaries, can be obtained by rotation in a magnetic field (Figure 4). The SANS experiment provides the important following qualitative information: when the particle volume fraction increases, the scattering signals show an increase of intensity and number of lamellar reflections. (Unfortunately, the resolution of the SANS setup did not allow exploiting the profile of these quasi-Bragg peaks.) Moreover, the featureless scattering signal, at smaller angles, simultaneously disappears. These observations completely disagree with the Helfrich model and strongly suggest that the Helfrich repulsive interaction between membranes is deeply altered by the insertion of the maghemite particles. Furthermore, these features point to a strong increase of the elastic compression modulus (at constant chemical potential $B^*$), upon doping. Note that quite similar observations were made when the lamellar phase of the non-ionic surfactant $C_{12}E_5$ was doped with latex particles.[35]

Another investigation of the phase diagram and the structure, by SAXS, of doped lamellar phases shows that this conclusion is also valid for non-magnetic particles.[36] Indeed, the same lyotropic lamellar phase was swollen with a colloidal suspension of non-magnetic iron oxide particles in cyclohexane. The main features of the phase diagram of this system are the same as that involving the magnetic particles. For example, a minimum swelling is required for incorporation of rigid particles ($d_o > D$), as observed quite generally with mineral particles.[14] This study suggests that magnetic interactions do not influence the way in which the particles alter the Helfrich interactions.

In another study, the role of the membrane interactions was investigated in a different way.[37] First, an aqueous ferrofluidic suspension of negatively-charged, citrate-covered maghemite particles was prepared. Then, the $L_\alpha$ phase of the SDS surfactant and pentanol co-surfactant was swollen with aqueous ferrofluidic suspensions of variable salinity. This procedure allows tuning the respective strengths of electrostatic repulsions that are dominant when no salt is added and of Helfrich entropic repulsions that are dominant when large salt concentrations are used. There are two contributions to the membrane flexibility constant: $\kappa = \kappa_l + \delta\kappa_{elec}$, where $\kappa_l$ is the intrinsic (steric) constant of order $kT$ and $\delta\kappa_{elec}$, of order a few $kT$, is the contribution due to the electrostatic charges. In all cases, as expected, a minimum value, $d_{min}$, of the smectic period was found below which no particles can be incorporated into the $L_\alpha$ phase. However, it was found that $d_{min}$ depends on the salt concentration as the insertion of negatively-charged particles between two negatively-charged membranes has some energy cost. Altogether, particle doping is easier in $L_\alpha$ phases stabilized by Helfrich-type entropic repulsions than by electrostatic repulsions.

More information about the membrane flexibility $\kappa$ in Helfrich-type $L_\alpha$ phases can be obtained in a different way. Indeed, the swelling law that relates smectic periodicity to membrane volume fraction, $\phi_{mem}$, and takes the simple form $d = \delta/\phi_{mem}$ for rigid membranes where $\delta$ is the membrane thickness, shows a logarithmic correction for highly flexible membranes. Therefore, accurate measurements of d can give access to $\kappa$, provided that $\delta$ is also precisely known. This latter parameter can be obtained from a Porod plot of the small-angle scattering signal. This kind of investigation was performed with ferrosmectic samples and showed that $\kappa$ is not much affected by particle doping.[38] However, the presence of the maghemite particles seems to constrain the surfactant membranes by creating a sublayer from which they are excluded. The thickness of this sublayer increases from zero to about 1 nm when $\phi$ increases from zero to 4%. This interpretation was confirmed by a detailed analysis of optical textures.[39] Dynamic studies described in a next section have shown that this sublayer actually represents a fraction of particles adsorbed on the membranes.

The bending elastic constant, $K = \kappa/d$, and the modulus $B^*$, are two very important properties of a lyotropic lamellar phase that control its mechanical behavior through their combinations: the penetration length $\lambda^2 = K/B^*$ and the Caillé exponent $\eta = (\pi/2) kT d^{-2} (K/B^*)^{-1/2}$.[40] The name of $\lambda$ comes from the fact that a sinusoidal modulation with vector $q$, due for instance to substrate corrugation, decays exponentially within the bulk lamellar phase over a distance $1/(\lambda q^2)$.

$\lambda$ can conveniently be measured by the method of the Cano wedge.[40] The sample, in homeotropic anchoring conditions, is held between two glass plates that make a small angle. In the setup used with ferrosmectics, the sample was simply confined between a convex lens and a glass plate, which ensures a cylindrical symmetry.[41] In order to accommodate the thickness variation of the sample, a distribution of edge-dislocations appears and the positions of these lines-defects can be measured. From this data, through appropriate modeling, the dependence of $\lambda$ with particle volume fraction $\phi$ could be inferred. Two regimes could be distinguished: at low $\phi$ (< 1-2 %), $\lambda$ shows little dependence on $\phi$, meaning that $K$ and $B^*$ increase with $\phi$ in the same way; at high $\phi$ (> 1-2 %) however, $\lambda$ decreases with increasing $\phi$, meaning that $B^*$ has a stronger dependence than $K$ on $\phi$.

*Gold*

Small (around 2 nm in diameter) and hydrophobic gold particles were also inserted in direct or inverse columnar phases[26b] or in lamellar phases where the membranes were alkane-swollen.[42] In the latter study, the particles were shown to enter the membranes only above a critical hydrophobic thickness, opening the perspective of using these media as size filters.

**Membrane inclusions**

A special case is that of hydrophobic or amphiphilic particles in solvent-free membranes. The conditions on particle size and surface properties are more stringent than for swollen bilayers, but they have the notable advantage of giving access to unperturbed membranes, in their ‰natural+state. The particles can then act as probes of these two-dimensional complex media, with possible applications in the study of biomembranes, as discussed below.

*Membrane proteins and antimicrobial peptides*

First among these inclusions are of course biomolecules (such as integral membrane proteins and antimicrobial peptides) that associate with the cell membrane. Since this field of study is immense, we shall only overview a very specific topic, namely measuring the membrane-mediated interaction between inclusions. To date, very few experiments attempted to measure this interaction directly. First among them, freeze-fracture transmission electron microscopy (FFEM) studies yielded the radial distribution function of the inclusions.[43] Comparing the data to liquid state theories[44] resulted in a hard-core model with, in some cases, an additional repulsive or attractive interaction. FFEM was not extensively used, undoubtedly due to the inherent experimental difficulties; moreover, it is not obvious that the distribution measured in the frozen sample is identical to that at thermal equilibrium.

The interaction between membrane inclusions can also be studied using SAXS and SANS on oriented samples, an approach pioneered by Huang and collaborators.[45] These techniques are particularly well adapted to the problem at hand due to their non-invasive character and to the wavelength being close to the typical length scales to be probed. One can thus measure the structure factor of the two-dimensional system formed by the inclusions in the membrane and obtain the interaction potential between them.

This strategy was recently used to study alamethicin pores in DMPC membranes[46] and gramicidin pores in several types of membranes.[47]

*Hybrid inclusions*

Another class of potential membrane probes is that of custom-designed hybrid particles, consisting of an inorganic core functionalized by various ligands. For instance, tin oxide clusters functionalized with butyl chains[48] were inserted within bilayers formed by the zwitterionic surfactant DDAO and their interaction was quantitatively determined.[49]

*Quantum dots*

We have already mentioned that hydrophilic QDs were inserted in the water medium of lamellar phases. Owing to their small size, these fluorescent particles (with an appropriate hydrophobic functionalization) are good candidates for insertion within lipid or surfactant membranes.[50] It is noteworthy that the QDs were inserted within the (solvent-free) membrane of giant vesicles filled with magnetic nanoparticles, which can potentially be used as contrast agents for MRI and optical imaging techniques.

**Nanorods**

*Iron oxide nanorods*

Inorganic particles are usually more polydisperse than their biological counterparts, but they also have some advantages: higher contrast in microscopy or scattering techniques, strong optical properties and sensitivity to external fields (electric or magnetic). Furthermore, they can potentially yield new materials with finely-tuned properties. For instance, water dispersions of goethite ($\alpha$-FeOOH) nanorods are known to form isotropic, nematic and columnar phases with unusual magnetic properties. Inserting these particles in a dilute lamellar phase of nonionic surfactant yields a hybrid mesophase, easily aligned by thermal annealing and responsive to applied magnetic fields.[14] At high particle concentration (8 vol%) the nanorods acquire a two-dimensional nematic order, all the while remaining homogeneously dispersed within the lamellar matrix.[51] The latter becomes stiffer as the particle concentration increases, showing the intimate coupling between the two components.

*Gold nanorods*

As we have seen above, host mesophases can impart their positional and orientational order to the inclusions. This feature is of particular interest in the case of gold nanorods, which exhibit anisotropic optical properties, as discussed in section **Electronic properties**.

Indeed, one can envision the bottom-up formulation of optical metamaterials,[52] whose properties can be finely tuned by controlling the properties of the matrix. A first step in this direction was taken by Liu et al.,[53] who dispersed gold nanorods in nematic, cholesteric and columnar hexagonal lyotropic phases and showed that the the nanorods acquire a positive or negative order parameter, depending on the nature of the phase.

**Nanosheets**

*Clays*

The insertion of nanosheets into lyotropic lamellar phases was achieved more than ten years ago by using Laponite RD, a synthetic clay of the hectorite type, manufactured by Laporte Company. These nanosheets have an average diameter of $D \sim 30$ nm and a thickness of $t = 1$ nm, leading to a specific area of 800 m$^2$/g. The particle diameter is quite polydisperse but their thickness is well defined. These particles bear a negative electrical charge density. Aqueous suspensions of laponite particles must be kept at pH > 9 in order to avoid congruent dissolution.

In an initial study, laponite particles were inserted in the negatively charged lamellar phase of AOT (sodium bis (2-ethylhexyl)sulfosuccinate), that appears at surfactant mass fractions $w$ above 10% and up to $w \sim 60\%$. This system was studied by adsorption and osmotic pressure measurements and by SAXS. Although AOT adsorbs much less on laponite than cationic surfactants, it nevertheless forms, at pH 8, a complete bilayer on the particle rim. Very little laponite could be incorporated in the lamellar phase of AOT: the maximum mass fraction of laponite increased from 0.01% at $w = 20\%$ to 0.7% at $w = 60\%$.[54] Any further increase of doping led to a microscopic phase separation in a laponite-rich phase and a surfactant-rich one. In the single $L_\alpha$ phase domain, where the thickness, $d_w$, of the water sublayers between the surfactant membranes is larger than the nanosheet thickness, the laponite particles insert into the aqueous medium. This behavior is highly reminiscent of that of the spherical hydrophilic inclusions described above, like POMs, whose insertion can only be achieved if $d_w > D$. The SAXS study did not evidence any influence of laponite doping on the Caillé exponent $\eta$. For $d_w < t$, the coexistence of two distinct lamellar phases suggests that some particles may be inserted within the surfactant membranes.

The insertion of the same laponite nanosheets into the swollen lamellar phase of non-ionic $C_{12}E_4$ and $C_{12}E_5$ surfactants was also achieved by the same authors.[55] In a first step, an adsorption study was performed because a significant proportion of the surfactant may adsorb onto such highly anisotropic mineral particles. Indeed, the ethylene oxide head of these non-ionic surfactants adsorb onto laponite and form a surfactant layer at the particle surface, with a mean area of 0.47 nm$^2$ per surfactant. However, this amount of adsorbed surfactant still remains negligible compared to the total surfactant content in these systems. A maximum of 2 % (resp. 0.6 %) of laponite could be inserted in the water region of the $L_\alpha$ phase of $C_{12}E_5$ (resp. $C_{12}E_4$). There again, more doping resulted in microscopic phase segregation. The amount of inserted laponite followed a non-monotonic dependence on surfactant volume fraction as it first increased and then decreased with surfactant concentration. This complex behavior was explained in the following way: at high surfactant concentration, $d_w$ must be larger than the thickness of the particle covered by the surfactant brush whereas, at low surfactant concentration, the doped phase is destabilized because thermal undulation modes are quenched by the nanosheets. SANS experiments were performed to study this system in more detail as the scattered signal from either surfactant or laponite could be extinguished by contrast variation obtained by adjusting the D$_2$O/H$_2$O ratio. When the surfactant signal was absent, the scattering of the mixed system was similar to that of pure laponite suspensions, showing the good dispersion of the particles. When the laponite signal was extinguished, fitting of the lamellar reflections showed that laponite addition does not affect the Caillé exponent $\eta$.

## Dynamic properties

**Diffusion coefficients (FRAP, QELS, PFG-NMR)**

*Self-diffusion coefficients via FRAP*

Using fluorescence recovery after photobleaching (FRAP) one can determine the self-diffusion coefficients of fluorescent probes in an aligned lamellar or hexagonal phase, along the phase director and in the transverse direction, as demonstrated for small fluorescent molecules dissolved in the water or in the medium of the alkyl chains.[56] The diffusion coefficient of larger inclusions is sensitive to the mesoscopic structure of the phase (thickness of the bilayers and of the water layer) and has been used to test and refine hydrodynamic models for the particle mobility.[57]

*QELS experiments with ferrosmectics:*

Quasi-elastic light scattering (QELS) experiments were performed on isotropic ferrofluid samples and on aligned ferrosmectic samples.[58] With the ferrofluid, the time decay of the autocorrelation function that is not monoexponential due to polydispersity was fitted by a second order cumulant expansion.[59], with an average diffusion constant $<D> = D_0 \sim 1.5\times10^{-11}$ m$^2$.s$^{-1}$. This is in good agreement with the value predicted from the Stokes-Einstein relation.

With ferrosmectics, when the wave vector is parallel to the layers, two modes were detected that correspond either to particle concentration fluctuations or to layer displacement fluctuations ("baroclinic" mode). The first mode scatters polarized light whereas the second one scatters depolarized light, which allows separating them, in principle. Moreover, the first mode that arises from particle diffusion parallel to the layers, with a diffusion constant $D_\perp$, was found to be much stronger than the second. $D_\perp$ shows a linear dependence with $1/d_0$ ($d_0$ is the thickness of the cyclohexane sub-layer) that extrapolates to $D_0$ when $1/d_0$ goes to zero. At largest achievable confinement, $D_\perp$ decreases to about $10^{-11}$ m$^2$.s$^{-1}$. Upon confinement, $D_\perp$ is therefore reduced compared to $D_0$, but only by a small amount. This moderate decrease is at odds with the Faxén model that predicts how the hydrodynamic drag of a particle, and therefore its diffusion constant, is affected by the confinement between two rigid walls.[60] This discrepancy may arise from hydrodynamic slip at the confining walls that are not rigid planes but liquid-like assemblies of surfactants.

When the wave vector is almost perpendicular to the layers, the frequency of the scattered signal goes to zero, which is the sign of the absence of permeation through the layers on the time scale of these experiments.

A detailed and fairly comprehensive theoretical description of the elastic and hydrodynamic properties of doped lamellar phases was given in reference [61]. It aims at describing the QELS spectra of these mesophases and it is valid when the particles are located in the solvent sub-layers, irrespective of the nature (magnetic or not) of the particles. The theory predicts new fluctuation modes where particle concentration and layer displacement are coupled.

The interpretation of QELS data on ferrosmectics in the frame of this model shows that the elastic compression modulus $B^*$ increases from 38 Pa for the undoped phase to 230 Pa at a particle volume fraction of 0.8 % (in the hydrophobic medium). The splay elastic constant, $K$, of the lamellar phase increases more slowly, from $8\times10^{-14}$ N to $10^{-13}$ N.

*PFG-NMR of POM-doped $L_\alpha$ phases:*

Pulsed-field gradient nuclear magnetic resonance (PFG-NMR) is another way of measuring self-diffusion coefficients in lyotropic systems.[62] The POM-doped $L_\alpha$ phase of the Brij-30 system is particularly well-suited for such studies for the following reasons: (i) aligned samples can be produced by applying strong magnetic fields, which gives access to the anisotropy of diffusion; (ii) the phosphorus nucleus at the center of the POMs is very well shielded and therefore has a large $T_1$ relaxation time, which opens a large temporal window, (iii) D$_2$O can be used to formulate the system, which allows investigating the diffusion behavior of water and compare it with that of the POMs.

Both water and POM diffusion were found to display a very large anisotropy, with $D_\perp$ much larger than $D_{//}$, where $D_{//}$ is the diffusion constant along the normal to the layers whereas $D_\perp$ is the diffusion constant perpendicular to the normal (and parallel to the layers.)[63] Actually, the diffusion constants of both water and POMs, along the normal to the layers were found to be negligible ($D_\perp/D_{//} \sim 100$) on the time scale of these NMR experiments, which suggests that POM doping does not induce the formation of defects in the $L_\alpha$ phase.

The diffusion constant of water parallel to the layers does not depend on $\phi_{POM}$ but depends on the confinement. As expected, $D_\perp$ decreases with increasing confinement but it decreases much faster than predicted by the Faxén model. This was classically interpreted by considering an increasing proportion of water molecules "bound" to the PEG brushes through a hydrogen bond network, in fast exchange with "free" water molecules.

The diffusion constant of the POMs is almost two orders of magnitude smaller than that of free POMs (measured with a simple POM solution). Moreover, it slowly decreases upon increasing $\phi_{POM}$ and it is independent on the confinement. This latter observation is completely at odds with the Faxén model but it could be interpreted by the diffusion of the POMs within the PEG brushes of the membranes. This actually confirms the assumption given above that the POMs adsorb onto the PEG brushes of the surfactant membranes.

**Bulk rheology**

Large colloidal inclusions can change the rheological properties of mesophases by interacting with their textural defects.[64] The role of small particles, which are intimately mixed with the matrix has not been much investigated so far, with the exception of the two works that we describe below.

The POM-doped $L_\alpha$ phase that was discussed above has been the subject of rheological and in-situ SAXS (RheoSAXS) and in-situ Small-Angle Light-Scattering (RheoSALS) studies. Rheological measurements showed that POM-doping reduces the steady-state viscosity by as much as one order of magnitude and the yield stress by a factor of four.[65] The reduction in steady-state viscosity essentially depends on the fraction of surfactant membrane covered by the POM particles, regardless of the surfactant/water ratio. Such large effects of POM insertion on the macroscopic mechanical properties of the $L_\alpha$ phase were attributed to a reduction in topological defect density.

Moreover, the morphological transition from the $L_\alpha$ phase to the so-called "onion state" (i.e. the close-packed assembly of multi-lamellar vesicles, MLVs) that is classically observed with un-doped lamellar phases upon shearing is pushed to higher shear rates. In contrast, the shear stress at which the transition occurs does not depend on POM concentration and since the viscosity is reduced by POM insertion, then, a higher shear rate must be reached to produce onions. At high POM doping, the transition to the onion state could not even be reached at the highest shear rate experimentally accessible.

Finally, measurements by SAXS of the intensities of the lamellar reflections demonstrate that the POMs remain encapsulated within the MLV membranes, which shows that this technique is a simple and cheap way to incorporate inorganic particles within surfactant vesicles.

Note that onions containing ferrofluid particles about 5 nm in diameter could be produced by shearing the lamellar phase of lipids doped with aminosilane-functionalized maghemite nanoparticles.[66] The efficiency of the encapsulation of the ferrofluid particles almost reached 100% at low nanoparticle concentration but decreased down to ~ 30% at the maximum maghemite concentration used in this study (Fe$_{III}$ = 125 mM). In the future, such maghemite-doped MLVs may find biomedical applications such as contrast agents in Magnetic Resonance Imaging, for instance.

The formation of onions under shear was also observed in a $L_\alpha$ phase doped with the same laponite clay nanosheets described above.[67] A rheological investigation showed that the MLVs form at the same shear stress in doped and in pure lamellar phases. Moreover, the ($C_{12}E_4$, $D_2O$, laponite) mixed system was studied, under shearing conditions, by RheoSALS and SANS (RheoSANS) experiments where the laponite was contrast-matched. Starting from a stable doped lamellar phase, a strong change (up to ~ 30%) in lamellar spacing was observed and interpreted as the sign of a phase separation into clay-rich and surfactant-rich phases. In a given range of shear stress, two lamellar phases coexist just before complete phase separation occurs. No change in the peak positions was observed when shearing was stopped, which points to the irreversibility of these phenomena. These experimental results were tentatively explained in the following way: the cores of the MLVs, which are the regions of higher curvature, are depleted in stiff laponite nanosheets. This creates a water osmotic pressure difference between the clay-poor and clay-rich regions, leading to a water drain from the former to the latter, which may account for the large changes in lamellar spacing under shear, as soon as the onions are formed.

**Microrheology and electrophoresis**

Doping lamellar mesophases with minute amounts of nanometric spherical particles allows investigating the rheological properties of these systems at the microscopic level and over a frequency range extending to higher frequencies compared to traditional macroscopic rheology.[68] For example, latex particles of ~ 40 nm diameter were inserted into the aqueous space between the membranes of the swollen lamellar phase of the ($C_{12}E_5$, hexanol, water) system. By varying the membrane volume fraction, the lamellar period could be adjusted at will between 50 and 300 nm. The electrophoretic mobility of the confined probe particles was measured as a function of frequency by applying a sinusoidal electric field. This spectroscopic technique revealed the existence of two relaxation modes. A fast mode (~ $10^{-4}$ s) was interpreted as arising from the movement of particles trapped between collision points of the fluctuating membranes whereas a slower mode (~ $10^{-3}$ s) was related to the persistence length of the membranes.

# Electronic properties

**Noble metals Ë UV-Vis absorbance**

Due to the plasmon modes of the conduction electrons, noble metal nanoparticles exhibit striking optical properties (in particular, negative dielectric permittivity ε over a certain frequency range), which depend strongly on their size and shape,[69] and also on their relative position and orientation.[70] This makes them uniquely adapted for applications ranging from sensing[71] and imaging[72] to photonics.[73]

Owing to their symmetry, nanospheres have a single plasmon resonance, while rods have two, a transverse one (roughly at the position of the sphere resonance) and a longitudinal one (at larger wavelength), when the light polarisation is perpendicular and parallel to the rod length, respectively.

This microscopic anisotropy (at the particle level) is preserved at the macroscopic scale in aligned materials, as demonstrated recently for gold nanorods contained in nematic and hexagonal lyotropic phases.[53a] In this work, rods of moderate aspect ratio (the length-to-diameter ratio $L/D$ ~ 3) stabilized by the cationic surfactant CTAB were dispersed in micellar systems of the same surfactant in water, with benzyl alcohol as a co-surfactant. The anisotropic medium induces a strong particle alignment that can be measured directly via FFTEM and indirectly via the intensity of the longitudinal plasmon peak. The orientation of the phase (and hence of the rods) can be controlled by shear or by applying a strong magnetic field. Helical structures can also be obtained, by using a chiral additive.[53b]

One of the possible applications for noble metal particles dispersed in ordered media is the bottom-up assembly of metamaterials (materials structured over length scales below that of the radiation wavelength). Indeed, using as building blocks metal nanoparticles with an intrinsic negative permittivity can yield, for instance, materials with a negative index of refraction, as already demonstrated in top-down nano-fabricated materials.[74]

Self-assembly strategies for metamaterial synthesis would have a number of advantages:
- Yielding bulk materials of almost unlimited thickness, in contrast with conventional layer-by-layer approaches.
- Using pre-synthesized monocrystalline nanoparticles reduces the losses associated with crystal defects with respect to lithography fabrication.[75]

However, they also need to satisfy quite stringent conditions:
- Precise control over the position and orientation of the nanoparticles, in order to tune the coupling between individual plasmon modes[70] and thus the optical properties of the resulting structure.
- A high particle concentration (of about ten percent by volume, for rods with an aspect ratio of ten[74, 76]) is required for significant changes in the refractive index. This value is at least ten to a hundred times larger than in current equilibrium dispersions[53b] (although it can be reached in precipitated super-crystals).[77] At such concentrations, the Van der Waals attraction can become very important and must be balanced by appropriate repulsive interactions to preserve phase stability. In this context, the interaction between inclusions induced by the matrix can play an important role. It is encouraging to note that small and spherical gold particles inserted within a lamellar phase of nonionic surfactant acquire a repulsive interaction component that is stronger than the Van der Waals attraction.[42]

**Polyoxometalates - photochromism**

An obvious advantage of considering POMs as dopants for lyotropic lamellar phases lies in the POM spectroscopic properties. For example, the aqueous solutions of the commercial $PW_{12}O_{40}^{3-}$ POM are colorless but they turn blue when the POMs are reduced upon UV-irradiation. In a similar way, the POM-doped $L_\alpha$ phase is also photochromic as it turns blue upon UV-irradiation but it is much more sensitive to UV-light than the plain POM solution.[20] This enhanced photochromism is probably due to the oxidation of the alcoholic polar groups of the Brij-30 surfactants by the POMs.

**Iron oxides Ë magnetic properties**

*Ferronematics and ferrocolumnar phases*

To the best of our knowledge, the idea of imparting strong magnetic properties to nematic liquid crystals was first put forward to Brochard and De Gennes in 1970 by doping a nematic liquid crystal with magnetic particles.[78] This idea was soon realized[79] as ferronematics were produced and adding minute amounts of ferrofluids to nematic (or cholesteric) phases became popular in order to align them in magnetic fields and study them in more detail.[80] Afterwards, the ferrosmectics described in this article were elaborated. More recently, columnar phases doped with magnetic particles (ferrocolumnars) were also formulated and show physical properties more or less similar to ferrosmectics.[81] Nowadays, adding magnetic particles to isotropic or liquid-crystalline phases in order to induce or improve their alignment in a magnetic field is still being used.[82] In the following, we still focus on doped lamellar phases.

*Ferrosmectics:*

As mentioned previously, ferrosmectics are quite sensitive to small magnetic fields. Quantitative experiments were performed on homeotropically aligned samples held in optical flat glass capillaries of thickness $l$, submitted to a magnetic field parallel to the smectic axis, in order to measure the critical field, $H_c$, at which phase reorientation occurs.[83] $H_c$ scales as $l^{1/2}$, which is classical for smectics,[40] and as $\phi^{-1/4}$. No dependence of $H_c$ on the lamellar period $d$ was found. The mechanism of reorientation involves the nucleation of toric defects, in the mid-plane of the sample, which grow and proliferate until a fan-shaped texture is obtained. This is in contrast with the Helfrich-Hurault homogeneous instability that usually occurs in smectics. A model based on the balance of the elastic deformation energy due to the defect and of the magnetic energy was used to explain the data. It was further argued that the reorientation transition could be described in terms of a first order "phase" transition, i.e. with coexistence of planar and homeotropic regions in the sample.

A more direct approach involved magnetization measurements of ferrofluids and ferrosmectics performed either with Foner or with SQUID magnetometers.[84] Single-domain samples either in homeotropic or planar anchoring conditions were used. The susceptibility $\chi_0$ of the ferrofluid samples is proportional to the particle volume fraction: $\chi_0 = \alpha_0 \phi$, with $\alpha_0 = 13.3$. This dependence is quite consistent with the particle diameter and magnetization (including the polydispersity distribution). As expected, the susceptibility of the ferrosmectics is anisotropic, with $\chi_\perp < \chi_{//}$, where $\chi_\perp$ (resp. $\chi_{//}$) is the susceptibility in the direction perpendicular (resp. parallel) to the layers. The anisotropy of magnetic susceptibility reaches 13 % and both susceptibilities are proportional to the particle volume fraction: $\chi_{//} = \alpha_{//} \phi$ and $\chi_\perp = \alpha_\perp \phi$ with $\alpha_{//} = 11.8$ and $\alpha_\perp = 10.3$. A conclusion of this study is that the coupling mechanism between particles and membranes may arise from the existence of either ellipsoidal particles or pairs of particles whose reorientation is physically hindered by the surfactant membranes.

A study of the dynamics of the magneto-optic behavior brought confirmation of this assumption.[85] The field-induced birefringence of both ferrofluids and ferrosmectics was investigated by applying field step-functions to the samples. Even though maghemite has a cubic structure, ferrofluids still display field-induced birefringence mostly due to small shape and surface anisotropies. Upon field removal, this birefringence decays with a characteristic time of ~ 5 µs that corresponds to the Brownian orientational relaxation of the maghemite particles. In contrast, two characteristic times are observed with ferrosmectics. The first one is the fast relaxation observed with the free particles of ferrofluids whereas the second one is much slower (~ 15 ms). This slower mechanism was interpreted as due to a 10-20 % proportion of particles adsorbed on the membranes. Moreover, linear dichroism measurements show that these adsorbed particles are either ellipsoidal or even actually pairs of particles and that they tend to keep their magnetic moments parallel to the surfactant membranes. Note that this would indeed provide an explanation for the strong magnetic coupling between maghemite particles and surfactant membranes observed in ferrofluids.

*Lamellar phases doped with large spherical magnetic particles:*
Ferrosmectics are generally prepared by inserting into a lamellar phase magnetic spherical guest particles of diameter smaller than the thickness of the hydrophilic or hydrophobic host sublayers. Recently, a completely different strategy was proposed where the diameter of the magnetic particles is much larger than the period of the lamellar phase.[15, 82b] In this case, the particles aggregate within the cores of the topological defects of the liquid-crystalline phase. This provides a new way of aligning the mesophase in a magnetic field, as illustrated by doping the $L_\alpha$ phase of a monolinolein lipid system with magnetite $Fe_3O_4$ nanoparticles. At high temperature, in the isotropic liquid phase, and in zero-field, the magnetic particles are well dispersed but they form linear aggregates in moderate magnetic fields (~ 1T) aligned along the field direction. When the temperature is slowly decreased, the lamellar phase nucleates with preferential alignment along the field direction. Therefore, the mesophase alignment is due to its heterogeneous nucleation on the linear particle aggregates. This process is both reproducible and reversible since both the lamellar phase and the particle aggregates vanish when the sample is heated up in the isotropic fluid phase. Then, the magnetic field can be applied again, possibly in a different direction, and the sample can be cooled again. The efficiency of this indirect mechanism strongly depends on the particle concentration. Indeed, the alignment of the lamellar phase, expressed in terms of nematic order parameter, regularly increases from 0.2 at 0.1 % particle concentration to 0.6 at 5 % particle concentration. (Note that the same type of investigations was successfully performed with the hexagonal mesophase.)

Moreover, magnetite nanoparticles strongly absorb light and this dissipation gives rise to a temperature increase of the hybrid sample (photothermal affect). In a sample of typically 50 mg and 5 % particle concentration, the light of a microscope source can heat the sample by up to 20°C within 5 minutes. Then, the lamellar phase melts and all birefringence is lost.[86] Thus, such systems are both light- and magnetic-field-responsive and might therefore find biomedical applications.

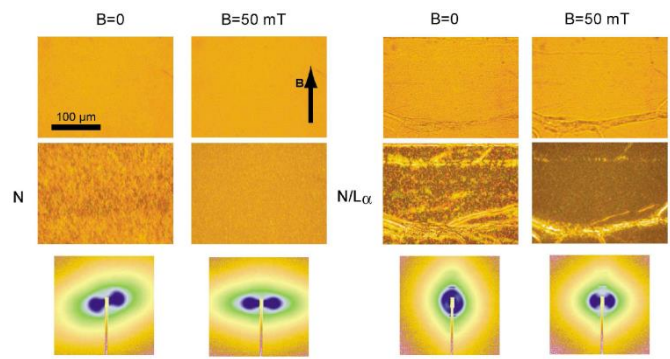

Figure 5: Optical microscopy textures (top) and SAXS signal (bottom) of the nematic phase of goethite, at a concentration $\phi_g = 8$ vol %, in water (left) and contained within the lamellar $L_\alpha$ phase (right). The lower microscopy images are taken between crossed polarizers, parallel to the image sides. In both cases, the nematic phase is very well aligned along the magnetic field, even at a relatively weak value of 50 mT. In the panel on the right, the lamellar phase is almost completely aligned in homeotropic anchoring (bilayers parallel to the flat faces of the capillary), with the exception of a few oily streak defects, visible in the microscopy images and which give rise to the very weak and sharp peaks along the vertical axis in the SAXS images. Reprinted with permission from reference [51].

*Lamellar phase doped with goethite nanorods:*

As mentioned above, goethite nanorods have very peculiar magnetic properties as they align parallel to the field at small field intensities (below 0.35 T) but reorient perpendicular to the field at large field intensities (above 0.35 T).[87] Magnetic measurements showed that this is due to a balance between their remanent magnetic moment and their negative anisotropy of magnetic susceptibility. When inserted in the aqueous sublayers of the $L_\alpha$ phase of the non-ionic surfactant $C_{12}EO_5$, goethite nanorods organize in an isotropic phase or in a nematic one, depending on their concentration, and keep their magnetic properties.[14] Consequently, although the undoped $L_\alpha$ phase hardly responds to strong magnetic fields, the small field delivered by permanent magnets allows reorienting the phase from homeotropic to planar orientation (Figure 5).

Here, the coupling mechanism between the nanorods and the membranes is clearly of steric origin because the particles are quite anisotropic, the direction of their magnetic moment is fixed with respect to their crystallographic structure, and they are strongly confined more or less parallel to the surfactant membranes.

Similar results were obtained when a lipidic lamellar phase was doped with amyloid fibrils coated with iron nanoparticles. Then, the mesophase could be aligned by slowly cooling the sample from its isotropic liquid phase in a 1T field.[88]

## Conclusion

This research field is growing fast, so any attempt at a summary would be premature. We can only offer some tentative conclusions:

- From a practical point of view, the change in the relative intensity of the Bragg peaks ensures that the particles are well dispersed in the matrix[26b] and can yield information as to their location.[20, 28] This analysis is convenient, because it can be done on laboratory instruments and does not require oriented samples; on the other hand, it works best with inorganic particles of high electron density.
- For a more thorough analysis, aligned samples are however indispensable if one wishes to measure independently the properties of the system along its symmetry axis and in the plane of the layers. This distinction is crucial for inherently anisotropic systems.
- A recurring theme is the importance of matching particle size and spacing of the host phase for successful doping.[14, 22, 42] Even for "soft" matrices, such as those considered here, the interaction engendered by (albeit small) deformations is sufficient to aggregate the particles.
- As dramatically illustrated by the observation above, the elastic coupling between inclusions and host phase is of paramount importance for the stability of the system and, surely, for many of its properties. Nevertheless, the influence of the inclusions on the elastic properties of the phase (and in particular on its fluctuations) is not yet well understood.
- Concerning the interactions between particles, one would like to know at the very least whether they are simply two-dimensional (within the same layer) or if there is significant coupling across the layers. Using very well aligned samples, this question could be answered for a few systems, with respect not only to positional[21, 49b] but also to orientational coupling.[89]

We are confident that the next few years will witness interesting developments along the lines sketched above.

## *Acknowledgements*

*We acknowledge funding from the Agence Nationale pour la Recherche under contracts FERROENERGY, MEMINT and NASTAROD.*

**Keywords:** liquid crystals · doping · inorganic materials · lamellar phase · nanoparticles

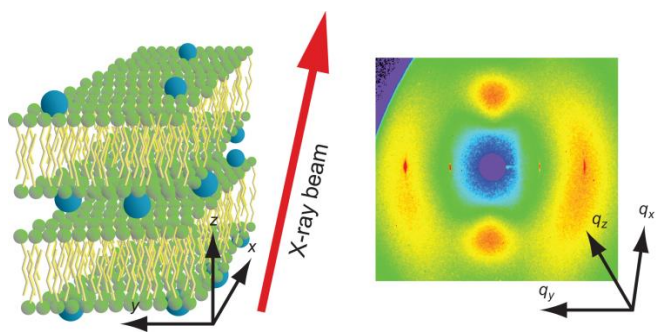

**Soft but ordered hybrids**. Inserting inorganic nanoparticles into soft lyotropic mesophases yields systems with new functional properties. Achieving macroscopic-scale orientation is essential for a complete characterization of these new materials, whether by structural, dynamical or spectroscopic techniques.